\documentclass[article,pra]{revtex4}
\usepackage{color}
\usepackage{xcolor}
\usepackage{graphicx}
\usepackage{bm}
\usepackage{amsmath}
\usepackage{amsbsy}
\usepackage{hyperref}
\usepackage{enumerate}
\usepackage{upgreek}
\usepackage[subnum]{cases}
\usepackage{dsfont}

\newcommand{\qm}{ \text{qm} }

\newcommand{\pa}{ \partial }
\newcommand{\hb}{ \hbar }

\newcommand{\ga}{ \gamma }
\newcommand{\la}{ \langle }
\newcommand{\ra}{ \rangle }

\newcommand{\Ga}{ \Gamma }

\newcommand{\Sch}{ \text{Sch} }

\newcommand{\traj}{\text{traj}}

\begin{document}

\title{ Trajectory-based Measure of Nonlocality in the Double Caldeira-Leggett Formalism }

\author{S. V. Mousavi}
\email{vmousavi@qom.ac.ir}
\affiliation{Department of Physics, University of Qom, Ghadir Blvd., Qom 371614-6611, Iran}
\begin{abstract}

We investigate the dynamics of quantum correlations in bipartite systems initially prepared in a squeezed state, comparing closed-system unitary evolution under the Schr\"odinger equation with open-system dynamics governed by the Caldeira-Leggett master equation in the high-temperature, weak-coupling regime, all within the Bohmian mechanics framework. Quantum nonlocality is quantified via the sensitivity of the Bohmian velocity of one particle to the position of the other. Our results show that in both distinct (local) and common bath scenarios, nonlocal correlations initially grow from zero, reach a peak, and then decay. In the case of local baths, the decay is smooth and monotonic; although the peak value increases with temperature, its temporal width (measured via the full width at half maximum) decreases, indicating a shorter duration of nonlocal correlations. For a common bath, the initial growth and decay are followed by revivals and oscillations, whose amplitude and timing vary with temperature. These non-monotonic behaviors arise despite the Markovian nature of the underlying dynamics and reflect the nontrivial role of system-bath correlations. We also analyze how both temperature and the squeezing decay parameter affect the structure of Bohmian trajectories and the evolution of nonlocal correlations. This trajectory-based, velocity-sensitivity measure offers an intuitive and quantitative understanding of entanglement degradation, decoherence, and their characteristic time scales. Our findings emphasize how the structure of the environment critically shapes the observable dynamics of quantum correlations, even in Markovian regimes.

\end{abstract}

\maketitle




\section{Introduction}

The unavoidable interaction of quantum systems with their environments gives rise to decoherence and dissipation, making the study of open quantum systems \cite{BrPe-book-2002, Ba-book-2018, Ri-book-2025} essential for understanding and developing realistic quantum technologies. A new framework models open-system dynamics via a multi-component quasi-stochastic process, yielding a super-quasi-cumulant expansion and a perturbative scheme for the dynamical map \cite{Sz-SPLN-2023}.
For open quantum systems weakly coupled to an environment near steady state, \cite{WaMi-bookchap-2025} outlines approaches ranging from master and Fokker-Planck equations to stochastic differential equations.

Among various approaches to modeling open-system dynamics, the Caldeira-Leggett (CL) formalism \cite{CaLe-PA-1983, Ca-book-2014} stands out for its physical transparency and analytical tractability. Originally developed to describe quantum Brownian motion, it models a system interacting linearly with a large reservoir of harmonic oscillators, leading to a reduced dynamics described by a master equation. In the high-temperature and weak-coupling limit, the CL master equation captures the essential features of dissipation and decoherence while remaining suitable for analytical treatment. 
Mousavi and Miret-Art\'es applied the CL formalism to study interference and diffraction in two-identical-particle systems, highlighting how dissipation and temperature affect decoherence and particle indistinguishability \cite{MoMi-EPJP-2020}. In a subsequent work, the same authors explored additional decoherence aspects, including arrival time distributions, diffraction in time, and the loss of symmetry in identical-particle systems, further emphasizing the role of temperature and relaxation rate in the quantum-to-classical transition \cite{MoMi-EPJP-2022}.
In the single-particle version, the formalism describes the dissipative evolution of a quantum particle subject to friction and diffusion due to its coupling with a thermal environment. The double CL formalism extends this framework to bipartite systems by coupling each particle either to independent environments or to a common bath \cite{CaMoPo-PA-2010}. The structure of the environment plays a fundamental role in the system dynamics: independent baths generally lead to Markovian decoherence characterized by monotonic decay of quantum coherence, whereas a common bath can mediate indirect interactions between particles, potentially introducing correlations and more complex dynamical behaviors. 

Quantum correlations are highly sensitive to environmental influence \cite{St-book-2015, FaPiAd-book-2017}. For a recent review of scalable methods to detect and characterize quantum correlations in many-body systems, see \cite{FrFaLe-RPP-2023}.
The degradation of quantum correlations is one of the central challenges in realizing entanglement-based technologies, such as quantum communication and computation \cite{BeBeFr-book-2023}. It is therefore important to understand how such correlations evolve under different environmental configurations and how they may be preserved or even revived under suitable conditions.
The constructive role of the environment in inducing quantum correlations, even when initially absent, has been examined in \cite{Mo-EPJP-2024} for a system of two coupled qubits described by an X-shaped state. 
The effects of decoherence and squeezing on the dynamics of quantum features have been studied within the CL formalism for systems initially prepared in squeezed states \cite{Mo-Arxiv-2025}. Such studies highlight that the environment can act not only as a source of noise but also as a resource in manipulating quantum correlations. 
The decoherence process in momentum space for cat states and identical spinless particles has also been investigated within this framework \cite{KhMoMi-Ent-2021}.
Using a variant of the CL model in which the environment is represented by a one-dimensional free scalar field, the saturation timescales of various quantum correlations have been investigated \cite{Bhetal-PRD-2023}.
It is notable that, in addition to conventional environmental decoherence, decoherence can be treated as intrinsic or modeled via quantum operations or decoherence channels. Reference \cite{Mo-QIP-2025} investigates the impact of these schemes on the dynamics of tripartite entanglement.

Among the various quantum states used to study correlation dynamics, the squeezed state occupies a prominent position. Squeezed states are characterized by reduced quantum uncertainty in one quadrature at the expense of increased uncertainty in the conjugate one, making them a key resource in quantum metrology \cite{Peetall-RMP-2018} and continuous-variable quantum information processing \cite{Laetal-PRX-2021}. 
In bipartite systems, two-mode squeezed states serve as standard models of entangled Gaussian states. They have been widely employed to investigate diverse aspects of quantum behavior. Lvovsky offered a comprehensive overview of squeezed light and its role in quantum optics \cite{Lv-SL-2015}. Adesso and Illuminati provided a seminal review of continuous-variable entanglement, emphasizing the foundational theory, conceptual structures, and mathematical methods for Gaussian states, as well as separability criteria, bipartite and multipartite entanglement properties, and the implications of the monogamy inequality. Their work also highlighted advances and open challenges in the characterization of entanglement in non-Gaussian states \cite{AdIl-JPA-2007}. More recently, van Loock and Shchukin discussed the entanglement properties of Gaussian states and analyzed the set of Gaussian operations relevant for their manipulation \cite{LoSh-QI-2016}. See also \cite{Se-book-2023} where the author provides a comprehensive modern overview of the principles and techniques of quantum continuous variables, from fundamental concepts to advanced applications in quantum information and technologies.

In this work, we analyze the nonlocal correlation dynamics and decoherence from the perspective of Bohmian Mechanics (BM) \cite{Bo-PR-1952, Tu-book-2021, Holland-book-1993}
, an alternative formulation of quantum theory in which a complete description of a quantum system involves both the wave function and the actual positions of particles, with the latter evolving under the guidance of the former. The minimal (first-order) formulation is based solely on the guidance equation, which dictates the trajectory of particles directly from the wave function, without introducing the concept of a quantum potential. In contrast, the second-order formulation, originally introduced by Bohm and often referred to as Bohm's dynamics \cite{CoVa-PRSA-2014} or quantum potential dynamics \cite{GoSt-JPA-2014}, replaces the guidance equation with a 
Newtonian-like equation of motion modified by an additional quantum potential term. 
To guarantee empirical agreement with standard quantum mechanics, the initial particle positions are postulated to follow a Born rule distribution. However, recent tunneling studies suggested contradictions with BM \cite{ShPuMaToKl-Nat-2025}, but Nikolić \cite{Ni-arXiv-2025} and Wang et al. \cite{WaWaWaLu-arXiv-2025} show these arise from misdefined velocities. Drezet et al. \cite{DrLaNa-arXiv-2025} further confirm BM's consistency, addressing the limitations of stationary treatments in the evanescent regime and the contribution of radiative leakage to the Bohmian velocity.
A novel double-slit experiment with a horizontal detection screen offers fresh tests of quantum interpretations \cite{AyKaBaGo-CP-2023}.
In entangled two-qubit systems, entanglement has been shown to modulate the degree of chaos in Bohmian trajectories through Lyapunov analysis \cite{TzCoZa-Ent-2025}, while the interplay of chaotic and ordered paths has been linked to the emergence of Born’s rule \cite{TzCo-PS-2021}. Complementary studies further demonstrate that near moving nodal points in rational frequency systems, the quantum potential can induce chaotic motion for specific wavefunction configurations \cite{UmNuChAh-PS-2025}.
Additionally, a connection between squeezed states and time-of-arrival measurements has been explored within the Bohmian mechanical framework \cite{GaLa-arXiv-2025}.

A central conceptual issue in BM is the treatment of the density matrix. While Bell famously argued that ``in the de Broglie-Bohm theory a fundamental significance is given to the wave function, and it cannot be transferred to the density matrix" \cite{Be-book-2004}, subsequent developments have challenged this view. By adopting an algebraic approach in which the density matrix is treated as a description of an individual system, it has been shown that the de Broglie-Bohm interpretation can be consistently extended to encompass density matrices \cite{Ma-FP-2005}.
Moreover, when spin or other internal degrees of freedom are considered, the density matrix plays multiple roles within BM. Beyond its standard interpretations as a statistical mixture or reduced state, it can also appear as a conditional density matrix, describing the state of a subsystem conditioned on another subsystem's configuration \cite{DuGoTuZa-FP-2005}.
Notably, expressing the density matrix in a complex-polar form enables the derivation of quantum equations of motion for Liouville-space trajectories, even in the presence of environmental dissipation \cite{MaBi-JCP-2001}. This broader interpretive framework aligns with the perspective advocated by Anandan and Aharonov \cite{AnAh-FPL-1999}, who argue that the density matrix should be regarded as a fundamental physical object rather than merely a statistical tool. In this spirit, we have employed a polar decomposition of the scaled von Neumann equation, which governs a smooth transition from quantum to classical dynamics \cite{MoMi-Sym-2023}.

Within this framework, it is possible to quantify nonlocal correlations using operationally meaningful quantities such as the sensitivity of the Bohmian velocity of one particle to the position of its entangled partner. This measure provides a direct trajectory-level insight into quantum nonlocality, complementing standard measures. In this paper, we investigate the dynamics of such nonlocal correlations in a bipartite system initialized in a two-mode squeezed state, and evolving under the double CL master equation. We consider both independent and common environmental couplings, and explore the roles played by temperature and the squeezing decay parameter. In addition to computing the nonlocality measure based on Bohmian velocities, we analyze the structure and behavior of Bohmian trajectories, thereby offering both intuitive and quantitative perspectives on the interplay between decoherence and correlation dynamics. Notably, inspired by Bohmian dynamics, two entanglement measures for pure states of bipartite quantum systems have been proposed and studied in the context of a nonlinear Schrödinger equation \cite{ZaPl-Ent-2018}. 

The rest of the paper is organized as follows. In Section \ref{sec: BM_CL}, we examine the double Caldeira-Leggett master equation within the framework of BM and introduce a velocity-sensitivity measure to quantify nonlocal correlations. Section \ref{Squeez_evol} is devoted to the evolution of the squeezed state from a Bohmian perspective. Our numerical results and related discussions are presented in Section \ref{sec: Res_Dis}. Finally, Section \ref{sec: Sum_con} concludes the paper with a summary of the main findings.

\section{Bohmian Analysis of the Caldeira-Leggett Master Equation} \label{sec: BM_CL}

In the high-temperature regime, the dynamics of a quantum particle interacting with a thermal environment can be effectively described by the CL master equation. Based on the path integral formalism, Caldeira and Leggett \cite{CaLe-PA-1983, Ca-book-2014} derived a non-unitary evolution equation for the reduced density matrix of the system. Assuming that the system is free (i.e., not subject to an external potential), the reduced density matrix $ \rho(x, y, t) $ in the position representation evolves according to
\begin{eqnarray} \label{eq: CL_1p}
\frac{\pa \rho}{\pa t} &=& \left[ - \frac{\hb}{2 m i} \left( \frac{\pa^2}{\pa x^2} - \frac{\pa^2}{\pa y^2} \right) - \ga(x-y) \left( \frac{\pa}{\pa x} - \frac{\pa}{\pa y} \right) - \frac{D}{\hb^2}(x-y)^2 \right] \rho(x, y, t) ,
\end{eqnarray}
where $m$ is the mass of the particle, and $\ga$ denotes the relaxation (or damping) rate induced by the coupling to the environment. The temperature of the reservoir is denoted by $T$, and the diffusion coefficient by $D$,
\begin{eqnarray}
D &=& 2 m \ga k_B T
\end{eqnarray}
where $k_B$ being Boltzmann's constant. Eq. \eqref{eq: CL_1p} captures both the unitary and dissipative aspects of the dynamics. The first term represents coherent evolution due to the kinetic energy. The second term describes dissipation, resulting from energy exchange with the environment. The third term accounts for decoherence and leads to the suppression of off-diagonal elements of $ \rho(x, y, t) $ in position space. This form of the master equation is valid in the weak-coupling limit, under the high-temperature assumption $k_B T \gg \hb \ga$. It describes a Markovian, memoryless process, as the right-hand side does not depend on the history of the density matrix--that is, it is local in time. Furthermore, the superoperator in the equation is independent of both time and the initial preparation \cite{Sc-PR-2019}.
By transforming to the rotated coordinates $ R = (x+y)/2 $ and $ r = x- y $, the master Eq. \eqref{eq: CL_1p} can be recast in a form that resembles a continuity equation, thereby highlighting the conservation of probability. In these variables, the evolution equation becomes \cite{MoMi-EPJP-2022}
\begin{eqnarray} \label{eq: CLeq_Rr_1p}
\left[ \frac{\pa}{\pa t} + 2 \ga r \frac{\pa}{\pa r} + r^2 \frac{D}{\hb^2} \right] \rho(R, r, t) + \frac{\pa}{\pa R} j(R, r, t)  &=& 0 ,
\end{eqnarray}
where the quantity $ j(R, r, t) = - i \hb \pa_r \rho(R, r, t) / m $ serves as the probability current density matrix. 
Imposing the condition $r=0$, which corresponds to evaluating the density matrix along its diagonal, yields the standard continuity equation for the probability density
\begin{eqnarray} \label{eq: con_CL-1p}
\frac{\pa P(x, t)}{\pa t} + \frac{\pa J(x, t)}{\pa x}  &=& 0 ,
\end{eqnarray}
where $ P(x, t) = \rho(R=x, r=0, t) $ represents the probability density in position space, and $ J(x, t) = j(R=x, r=0, t) $ is interpreted as the associated probability current. This formulation makes explicit the role of the CL equation in preserving total probability, even in the presence of dissipative and decoherence terms arising from the system's interaction with its environment.

In Bohmian mechanics, an alternative formulation of quantum theory, the complete state of a system is described by both its wavefunction and the actual position of the particle. To ensure empirical equivalence with standard quantum mechanics, the initial positions of particles are assumed to be distributed according to the Born rule \cite{Tu-book-2021, Holland-book-1993}. In the minimalist formulation of the theory, known as Bohmian mechanics \cite{Tu-book-2021}, particle trajectories are obtained by integrating the guidance equation,
\begin{eqnarray} \label{eq: guidance}
\dot{x}(x, t) = \frac{J(x, t)}{P(x, t)}\bigg|_{x = X(t, X^{(0)})}
\end{eqnarray}
where $X^{(0)}$ denotes the initial position of the Bohmian particle, and $X(t, X^{(0)})$ represents its trajectory---that is, the particle's actual position at time $t$.

An alternative approach, initially proposed by Bohm in his seminal 1952 paper \cite{Bo-PR-1952} and known as the de Broglie-Bohm pilot-wave theory, involves expressing the system's state in polar form and substituting it into the system's evolution equation. By separating the resulting expression into its real and imaginary components, one obtains a generalized Hamilton-Jacobi equation and a continuity equation, respectively. From the phase function appearing in the polar decomposition, a momentum field is defined via its spatial gradient, leading to a Newtonian-like equation of motion for the particle aspect of the system. This equation naturally incorporates the so-called quantum potential energy.

We now consider an open two-particle system described within the CL framework. Two coupling scenarios can be examined: (i) each particle interacts with an independent environment, and (ii) both particles are coupled to a common environment. In the case of distinct environments, the two baths are assumed to be identical, characterized by the same relaxation rate and temperature. Throughout this work, the dynamics are assumed to be free, with no external potential acting on either particle. When the particles are coupled to a common environment, the corresponding two-particle CL master equation takes the form \cite{CaMoPo-PA-2010},
\begin{eqnarray} \label{eq: CL-common-1}
\frac{\pa}{\pa t}\rho(x_1, y_1; x_2, y_2; t) &=& \sum_{n=1}^2 \bigg[ 
+ i \frac{\hb}{2 m} \left( \pa_{x_n}^2 - \pa_{y_n}^2 \right) 
- \ga (x_n-y_n) \left( \pa_{x_n} - \pa_{y_n} \right) 
- \frac{D}{\hb^2} (x_n-y_n)^2
\bigg]\rho(x_1, y_1; x_2, y_2; t) 
\nonumber \\
&-&
\bigg[ \ga \sum_n \sum_{n' \neq n } (x_n - y_n) \left( \pa_{x_{n'}} - \pa_{y_{n'}} \right)
+ 2 \frac{D}{\hb^2} (x_1 - y_1)(x_2 - y_2) \bigg] \rho(x_1, y_1; x_2, y_2; t) .
\end{eqnarray}
Note that the second line of this equation pertains to the common environment scenario; that is, when particles interact with two separate environments, this second line is not present.
Substituting the polar form
\begin{eqnarray} \label{eq: pol-form}
\rho(x_1, y_1; x_2, y_2; t) &=& A(x_1, y_1; x_2, y_2; t) ~ \exp\left[ \frac{i}{\hb} S(x_1, y_1; x_2, y_2; t) \right]
\end{eqnarray}
into \eqref{eq: CL-common-1} and splitting the real and imaginary parts of the resulting equation yields respectively the Hamiltonian-Jacobi-like equation
\begin{eqnarray} \label{eq: HJ-mixed}
-\frac{\pa S}{\pa t} &=&
\sum_{n=1}^2 \bigg[ \frac{ (\pa_{x_n} S)^2 }{2m} - \frac{ (\pa_{y_n} S)^2 }{2m} + \ga (x_n - y_n) (\pa_{x_n}S - \pa_{y_n}S )  \bigg] + \ga \sum_n \sum_{n' \neq n } (x_n - y_n) \left( \pa_{x_{n'}} S - \pa_{y_{n'}} S \right) 
\nonumber \\
&+& Q(x_1, y_1; x_2, y_2; t) ,
\end{eqnarray}
for the phase, and the continuity equation
\begin{eqnarray} \label{eq: con-A}
& &\frac{\pa A}{\pa t} +  \sum_{n=1}^2 \bigg[ \frac{1}{m} (\pa_{x_n} A~ \pa_{x_n} S - \pa_{y_n} A~ \pa_{y_n} S ) 
+ \frac{1}{2m} A (\pa_{x_n}^2 S - \pa_{y_n}^2 S ) 
+\ga (x_n - y_n) (\pa_{x_n}A - \pa_{y_n}A ) + \frac{D}{\hb^2} (x_n - y_n)^2 A \bigg] 
\nonumber \\
& \qquad & + \ga \sum_n \sum_{n' \neq n } (x_n - y_n) \left( \pa_{x_{n'}} A - \pa_{y_{n'}} A \right)
+ 2 \frac{D}{\hb^2} (x_1 - y_1)(x_2 - y_2) A = 0 ,
\end{eqnarray}
for the amplitude, where
\begin{eqnarray} \label{eq: Qp-mixed}
Q(x_1, y_1; x_2, y_2; t) &=& - \frac{\hb^2}{2m} \frac{ \sum_{n=1}^2 ( \pa_{x_n}^2 - \pa_{y_n}^2 ) A(x_1, y_1; x_2, y_2; t) }{A(x_1, y_1; x_2, y_2; t)} 
\end{eqnarray}
is the quantum potential. As Eq. \eqref{eq: HJ-mixed} shows the decoherence term in the CL equation does not contribute to the evolution of the phase.
But, the continuity equation \eqref{eq: con-A} contains quadratic terms proportional to $(x_n-y_n)^2$ and $ (x_1-y_1)(x_2-y_2) $, which suppresses the off-diagonal elements of the density matrix---those far from the diagonal---while leaving the elements along the diagonal-axis unaffected. This reflects the process of decoherence, namely the dynamical emergence of a diagonal density matrix as a result of environmental interactions.

By defining momentum and velocity vector fields 
\begin{eqnarray}
\begin{cases}
\bm{p}_n = \left( \pa_{x_n} S, - \pa_{y_n} S  \right) \label{eq: mom_fields}  \\
\bm{v}_n = \frac{ \bm{p}_n }{ m } \label{eq: vel_fields}
\end{cases}
, \quad n=1, 2
\end{eqnarray}
and the ancillary vector field
\begin{eqnarray}
\bm{\Ga}_n &=& \left( \ga(x_n-y_n), -\ga(x_n-y_n)  \right),
\end{eqnarray}
the Hamilton-Jackobi-like equation can be rewritten compactly as
\begin{eqnarray} \label{eq: HJ-mixed-1}
-\frac{\pa S}{\pa t} &=&
\sum_{n=1}^2 \bigg[ \frac{ (\pa_{x_n}S)^2 - (\pa_{y_n}S)^2 }{2m} + \bm{\Ga}_n \cdot \bm{\nabla}_n S  \bigg] +  \sum_n \sum_{n' \neq n } \bm{\Ga}_n \cdot \bm{\nabla}_{n'} S  
+ Q(x_1, y_1; x_2, y_2; t)  ,
\end{eqnarray}
while the continuity equation recasts
\begin{equation} \label{eq: con-A1}
\frac{dA}{dt} +  \sum_{n=1}^2 \bigg[ \frac{1}{2} A \bm{\nabla}_n \cdot \bm{v}_n + \bm{\Ga}_n \cdot \bm{\nabla}_n A
+ \frac{D}{\hb^2} (x_n - y_n)^2 A + \sum_{n' \neq n } \bm{\Ga}_n \cdot \bm{\nabla}_{n'} A \bigg] +  2 \frac{D}{\hb^2} (x_1 - y_1)(x_2 - y_2) A = 0 ,
\end{equation}
where we have used
\begin{eqnarray} \label{eq: t-rate-config}
\frac{\pa}{\pa t} +  \sum_{n=1}^2 \bm{v}_n \cdot  \bm{\nabla}_n  &=& \frac{d}{dt},
\end{eqnarray}
which is the time rate of change with respect to a moving point in configuration space.

Note that the density operator is Hermitian, $ \rho^{\dag}(t) = \rho(t) $. Thus, for its elements in the position representation one has that 
\begin{eqnarray}
\rho(x_1, y_1; x_2, y_2; t) &=& \la x_1, x_2 | \rho(t) | y_1, y_2 \ra = \la y_1, y_2 | \rho(t) | x_1, x_2 \ra^* =
\rho^*(y_1, x_1; y_2, x_2; t) ,
\end{eqnarray}
from which one obtains
\begin{numcases}~
A(x_1, y_1; x_2, y_2; t) = A(y_1, x_1; y_2, x_2; t), \\
S(x_1, y_1; x_2, y_2; t) = -S(y_1, x_1; y_2, x_2; t), \label{eq: S_anti-sym}
\end{numcases}
for the amplitude and phase of the density matrix which shows amplitude (phase) is (anti)-symmetric under the replacements $ x_n \leftrightarrow y_n $.

By independently applying the operators $\partial_{x_1}$ and $\partial_{y_1}$ to Eq.~\eqref{eq: HJ-mixed-1}, utilizing relations \eqref{eq: mom_fields} and \eqref{eq: t-rate-config}, and rearranging the terms, we obtain
\begin{eqnarray} \label{eq: d_dt_mom1}
\frac{d}{dt} \bm{p}_1 &=& - \ga ( \pa_{x_1} - \pa_{y_1} ) S - \big( \sum_n \bm{\Ga}_n \cdot \bm{\nabla}_n \big) \bm{p}_1 
\mp \bm{\nabla}_1 Q \nonumber \\
& \qquad & ~ - \ga ( \pa_{x_2} - \pa_{y_2} ) S - \big( \sum_n \sum_{n' \neq n} \bm{\Ga}_n \cdot \bm{\nabla}_{n'} \big) \bm{p}_1 
\end{eqnarray}
for the time rate of change of the first particle's momentum field in configuration space. In the third term of the first line, the minus (plus) sign corresponds to the $x$ ($y$) component. The second line represents the contribution from a common environment; for distinct, independent environments, only the first line applies. 
The equation for the second particle has a similar structure, with the indices 1 and 2 swapped.

Note that only the $x$ component is relevant for our purposes. The Bohmian velocity field of the particle $i$, $v_{iB}$, is given by the diagonal elements of the $x$ component of the momentum field divided by mass:
\begin{eqnarray} \label{eq: veli_def}
v_{iB}(x_1, x_2, t) &=& \frac{1}{m} \frac{\pa S}{\pa x_i} \bigg | _{y_n=x_n}, \qquad i = 1, 2 .
\end{eqnarray}
The anti-symmetry property of the phase function \eqref{eq: S_anti-sym} implies that $\big( (\pa_{x_i} - \pa_{y_i}) S \big) \big|_{y_n=x_n} = 2 ( \pa_{x_i} S ) \big|_{y_n=x_n}$.
Combining these results, we finally obtain
\begin{eqnarray} \label{eq: Newton1}
m \frac{d}{dt} v_{1B}(x_1, x_2, t) &=& - 2 \ga v_{1B}(x_1, x_2, t) - \frac{\pa Q}{\pa {x_1}} \bigg | _{y_n=x_n} - 2 \ga v_{2B}(x_1, x_2, t) ,
\end{eqnarray}
which resembles a Newtonian-like equation for the motion of the first particle. The last term arises only in the case of a common environment. The equation of motion for the second particle takes the same form, with the indices 1 and 2 interchanged.

By integrating the velocity fields given in Eq.~\eqref{eq: veli_def} and imposing appropriate initial conditions, one obtains the Bohmian trajectories $X_n(t, X_1^{(0)}, X_2^{(0)}) $. These trajectories exhibit explicit nonlocality: the velocity of one particle depends instantaneously on the actual position of the other.

Quantifying nonlocality within BM involves subtle conceptual and practical challenges. Although the theory is explicitly nonlocal and deterministic, its empirical predictions are fully aligned with those of standard quantum mechanics, where nonlocality typically emerges through statistical violations of Bell inequalities. In BM, by contrast, nonlocality is built directly into the dynamics via the quantum potential and the guidance equation. As a result, while direct experimental detection is constrained by this empirical equivalence, indirect approaches---such as those employing weak measurement techniques---have provided important insights. For instance, Wiseman \cite{Wi-NJP-2007} established a foundation for BM using weak values and Bayesian reasoning. Building on this, Kocsis et al. \cite{Koetal-Sc-2011} measured the average trajectories of single photons in a two-slit interferometer. 

To quantify the degree of nonlocality in Bohmian dynamics, we define a trajectory-based sensitivity measure that captures the dependence of the Bohmian velocity of particle 1 on the position of particle 2. For a specific trajectory initialized at $ ( X_1^{(0)}, X_2^{(0)} ) $, we introduce
\begin{eqnarray}
\eta_{\traj}(t; X_1^{(0)}, X_2^{(0)} ) &=& | \pa_{x_2} v_{1B} | \bigg|_{x_1 = X_1(t, X_1^{(0)}, X_2^{(0)}), x_2 = X_2(t, X_1^{(0)}, X_2^{(0)})}
\end{eqnarray}
as a indicator of the nonlocal effect that particle 2 exerts on the velocity of particle 1 along the trajectory, where the absolute value ensures positivity.
%
%
Averaging over all initial conditions weighted by the quantum equilibrium distribution yields the expected sensitivity
\begin{eqnarray} \label{eq: eta}
\eta(t) &=& \int \int dX_1^{(0)} dX_2^{(0)} ~ P_0(X_1^{(0)}, X_2^{(0)}) ~ \eta_{\traj}(t; X_1^{(0)}, X_2^{(0)} ).
\end{eqnarray}
Here, $ P_0(X_1^{(0)}, X_2^{(0)}) $ denotes the initial distribution of actual particle positions, determined by the Born rule. For pure states in the Schr\"odinger framework, this takes the form $ P_0 = | \Psi_0(X_1^{(0)}, X_2^{(0)}) |^2 $, while for mixed states it is given by the diagonal elements of the initial density matrix $ P_0 = \rho(X_1^{(0)}, X_1^{(0)}; X_2^{(0)}, X_2^{(0)}; 0) $.

At the end of this section, it is worth commenting on the conditional wave function (CWF), a useful concept unique to BM \cite{No-JPCs-2016}. The CWF of particle 1, given that particle 2 is at its actual position $X_2(t, X_1^{(0)}, X_2^{(0)})$, is defined as
\begin{eqnarray}
\psi_1(x_1, t) &=& \Psi(x_1, X_2(t, X_1^{(0)}, X_2^{(0)}), t) .
\end{eqnarray}
This is a function of the coordinates of particle 1 only, but it is parametrized by the actual position of particle 2. The guiding equation for particle 1 can be expressed entirely in terms of its CWF, taking the same form as the single-particle guiding equation. However, the CWF does not obey a standard Schr\"odinger equation. Its evolution depends on both the total wave function $\Psi$ and the motion of the other particle $X_2(t)$. This results in a modified, nonlinear, and non-unitary evolution equation for $\psi_1(x_1, t)$. Any motion of particle 2---or measurements performed on it---can cause an instantaneous, nonlocal change in the shape and phase of the CWF of particle 1. In this way, a remote action on particle 2 affects $\psi_1(x_1, t)$, thereby altering the quantum potential and velocity of particle 1.

Additional measures of nonlocality specific to BM could also be introduced--for instance, one based on the sensitivity of the CWF, such as $ || \pa_{X_2} \psi_1(x_1, t) || / || \psi_1(x_1, t) || $, where $ || f(x_1) ||^2 =\int_{-\infty}^{\infty} |f(x_1)|^2 dx_1 $ and $ \pa_{X_2} \psi_1(x_1, t) = \pa_{x_2} \Psi(x_1, x_2, t) \big |_{x_2 = X_2(t, X_1^{(0)}, X_2^{(0)}), t)} $. Another possibility is the weighted quantum force, defined as $ \int \int \big|  \pa_{x_1}\pa_{x_2 } Q(x_1, x_2, t)   \big|  | \psi(x_1, x_2, t) |^2 dx_1 dx_2 $, which quantifies how the quantum force acting on one particle varies with the position of the other.

\section{Bohmian Trajectory Dynamics of Squeezed States: Unitary and Dissipative Frameworks} \label{Squeez_evol}

In this section we analyze the time evolution of the bipartite squeezed Gaussian state
\begin{eqnarray} \label{eq: squ0_pos}
\Psi_0(x_1, x_2) &=& \frac{1}{\sqrt{\pi} } \exp \left[ - e^{-2s} \frac{(x_1+x_2)^2}{4}  - e^{2s} \frac{(x_1-x_2)^2}{4} \right], 
\end{eqnarray}
within the Bohmian formulation. Two dynamical regimes are considered: (i) unitary evolution governed by the Schr\"odinger equation and (ii) dissipative evolution described by the two-particle CL master equation.

Throughout the discussion we employ a dimensionless scheme in which the position variables $ x_n $ are already adimensional, and we subsequently set $ m = 1 $ and $ \hb = 1 $. With this choice every physical quantity---positions, times, velocities, momenta, forces, energies, and accelerations---appears in dimensionless form.

In Eq. \eqref{eq: squ0_pos}, the parameter $s$ characterizes the degree of squeezing. The spatial width of the state along the relative coordinate $ x_1 - x_2 $, quantified by the standard deviation $ \sqrt{ \la (x_1-x_2)^2 \ra - \la (x_1-x_2) \ra^2 } $, is given by $e^{-s}$. In the limit $ s \to 0 $, the state becomes separable, reducing to a product of two Gaussian wave packets centered at the origin, each with width $ 1 / \sqrt{2} $ and zero momentum. Conversely, in the limit $s \to \infty$, the state approaches the idealized form $\delta(x_1 - x_2)$, originally introduced by Einstein, Podolsky, and Rosen in their seminal critique of the completeness of quantum mechanical descriptions of physical reality \cite{EiPoRo-PR-1935}. 

For later convenience and to simplify expressions, we define
\begin{eqnarray} \label{eq: mu}
\mu &=& e^{-2s}
\end{eqnarray}
as a dimensionless parameter that characterizes the variance of the relative position distribution. Since $\mu$ decreases monotonically with increasing squeezing, we refer to it as the ``squeezing decay factor".

\subsection{Unitary Evolution under the Schr\"odinger Equation}

In the Schr\"odinger framework, the time-evolved wavefunction is given by
\begin{eqnarray}
\Psi(x_1, x_2, t) &=& \la x_1, x_2 | U(t) | \Psi_0 \ra = \int_{-\infty}^{\infty} dx_1 \int_{-\infty}^{\infty} dx_2 ~
G(x_1, x_2; x'_1, x'_2; t) ~ \Psi_0(x'_1, x'_2) ,
\end{eqnarray}
where $ G(x_1, x_2; x'_1, x'_2; t) $ denotes the two-particle propagator, i.e., the position-space matrix element of the unitary evolution operator $ U(t) = e^{- i H t / \hb} $, and $ | \Psi_0 \ra $ is the initial state. For free evolution governed by the Schr\"odinger equation, the propagator takes the form
\begin{eqnarray} \label{eq: propagator}
G(x_1, x_2; x'_1, x'_2; t) &=& \frac{1}{ 2 \pi  i t } \exp \left[ \frac{i (x_1-x'_1)^2}{2 t} + \frac{i (x_2-x'_2)^2}{2 t} \right],
\end{eqnarray}
which yields the following explicit expression for the time-evolved wavefunction
\begin{eqnarray} \label{eq: Sch-sol}
\Psi^{\Sch}(x_1, x_2, t) &=& \sqrt{ \frac{\mu}{ \pi( \mu + i \left(\mu ^2+1\right) t - \mu  t^2 ) } }
\exp \left[ 
\frac{2 i \mu  t \left(x_1^2+x_2^2\right)+\mu ^2 (x_1+x_2)^2+(x_1-x_2)^2}{4 (t-i \mu ) (\mu  t-i)}
 \right] ,
\end{eqnarray}
with the superscript ``Sch" indicating Schr\"odinger, as used here and in what follows.
Then, Bohmian velocity fields are given by,
\begin{numcases}~ 
v^{\Sch}_{1B}(x_1, x_2, t) = \frac{t \left(2 \mu ^2 t^2 x_1+\mu ^4 (x_1+x_2)+x_1-x_2\right)}{2 \left(t^2 + \mu^2 \right) \left(\mu ^2 t^2+1\right)} , \label{eq: BM_vel_1_Sch}
\\
v^{\Sch}_{2B}(x_1, x_2, t) = v^{\Sch}_{1B}(x_2, x_1, t) . \label{eq: BM_vel_2_Sch}
\end{numcases}
It is worth noting that in the absence of squeezing, these results reduce to the well-known expression $ v^{\Sch}_{nB} = x_n ~ t / (1+t^2) $ \cite{Holland-book-1993}.
Finally, Bohmian trajectories are given respectively by
\begin{numcases}~
X_1^{\Sch}(t, X_1^{(0)}, X_2^{(0)}) = \frac{\sqrt{t^2 + \mu^2} (X_1^{(0)}-X_2^{(0)})+\mu  \sqrt{\mu ^2 t^2+1} (X_1^{(0)}+X_2^{(0)})}{2 \mu } , \label{eq: BM_traj_1_Sch} \\
X^{\Sch}_2(t, X_1^{(0)}, X_2^{(0)}) = X^{\Sch}_1(t, X_2^{(0)}, X_1^{(0)}) \label{eq: BM_traj_2_Sch}.
\end{numcases}
As expected, for the special case $ \mu=0 $, corresponding to an unsqueezed initial state, the trajectory of each particle becomes independent of the actual position of the other.

Substituting Eqs. \eqref{eq: BM_traj_1_Sch} and \eqref{eq: BM_traj_2_Sch} into Eqs. \eqref{eq: BM_vel_1_Sch} and \eqref{eq: BM_vel_2_Sch} yields 
\begin{numcases}~
v^{\Sch}_{1B}(t, X_1^{(0)}, X_2^{(0)}) = \frac{t}{2\mu} \left( \frac{ X_1^{(0)} - X_2^{(0)} }{ \sqrt{ t^2 + \mu^2 } } + \mu^3 \frac{ X_1^{(0)} + X_2^{(0)} }{ \sqrt{ 1 + \mu^2 t^2} } \right), \label{eq: v1_sch_along}
\\
v^{\Sch}_{2B}(t, X_1^{(0)}, X_2^{(0)}) = v^{\Sch}_{1B}(t, X_2^{(0)}, X_1^{(0)}) , \label{eq: v2_sch_along}
\end{numcases}
for the velocities along a specific trajectory.

The quantum potential is obtained as
\begin{eqnarray} \label{eq: Qp_Sch}
Q^{\Sch}(x_1, x_2, t) &=& \frac{1}{4} \left[ 
2 \mu  \left( \frac{1}{\mu ^2 t^2+1}+\frac{1}{\mu ^2+t^2} \right) - \mu^2 \left( \frac{ (x_1+x_2)^2}{\left(\mu ^2 t^2+1\right)^2}+\frac{ (x_1-x_2)^2 }{\left(\mu ^2+t^2\right)^2} \right) \right],
\end{eqnarray}
which yields 
\begin{eqnarray} 
F^{\Sch}_{\qm, 1}(x_1, x_2, t) &=& \frac{ \mu^2 }{2} \left( \frac{x_1-x_2}{\left(\mu^2+t^2\right)^2}+\frac{x_1+x_2}{\left(\mu^2 t^2+1\right)^2} \right) , \label{eq: QM_force_1_Sch} \\
F^{\Sch}_{\qm, 2}(x_1, x_2, t) &=& F^{\Sch}_{\qm, 1}(x_2, x_1, t) , \label{eq: QM_force_2_Sch}
\end{eqnarray}
for the quantum forces acting on particles 1 and 2 respectively.
These quantum forces along trajectories reduce to
\begin{numcases}~
F^{\Sch}_{\qm, 1}(t, X_1^{(0)}, X_2^{(0)}) = \frac{\mu}{2} \left( \frac{ X_1^{(0)} - X_2^{(0)} }{ ( t^2 + \mu^2 )^{3/2} } + \mu \frac{ X_1^{(0)} + X_2^{(0)} }{ ( 1 + \mu^2 t^2 )^{3/2} } \right) , \label{eq: F1_sch_along}
\\
F^{\Sch}_{\qm, 2}(t, X_1^{(0)}, X_2^{(0)}) = F^{\Sch}_{\qm, 1}(t, X_2^{(0)}, X_1^{(0)}) . \label{eq: F2_sch_along}
\end{numcases}

By substituting \eqref{eq: BM_vel_1_Sch} in \eqref{eq: eta} and and applying the normalizability condition of the wave function, one obtains 
\begin{eqnarray} \label{eq: eta_Sch}
\eta^{\Sch}(t) &=& \frac{\left( 1 - \mu^4 \right) t}{2 \left( 1 + \mu^2 t^2\right) \left( t^2 + \mu^2 \right)}
\end{eqnarray}
for the nonlocality measure in the Schr\"odinger framework which represents a maximum at time 
\begin{eqnarray}
t^{\Sch}_{\max} &=& \sqrt{\frac{-\mu ^4+\sqrt{\mu ^8+14 \mu ^4+1}-1}{6 \mu ^2}},
\end{eqnarray}
with the maximum value
\begin{eqnarray} \label{eq: eta_Sch_max}
\eta^{\Sch}_{\max} &=& -\frac{3 \sqrt{6} \mu  \left(\mu ^4-1\right) \sqrt{-\mu ^4+\sqrt{\mu ^8+14 \mu ^4+1}-1}}{\left(-\mu ^4+\sqrt{\mu ^8+14 \mu ^4+1}+5\right) \left(5 \mu ^4+\sqrt{\mu ^8+14 \mu ^4+1}-1\right)}.
\end{eqnarray}

\subsection{Dissipative Evolution: Velocities and Equations of Motions}

The solution to the double CL equation with the initial state given in Eq.~\eqref{eq: squ0_pos} has been provided in Ref. \cite{Mo-Arxiv-2025} for both distinct and common environment scenarios, with the aim of analyzing the impact of decoherence on various quantum features and correlations. However, due to the length of the explicit expressions, we do not reproduce them here. Based on those solutions, the corresponding Bohmian velocity fields in the common environment case are obtained as
\begin{eqnarray} 
v_1^c(x_1, x_2, t) &=&
\frac{ f(x_1, x_2, t) }{2 \left(\mu ^2+t^2\right) \left(\mu  (\ga  \mu -D)+e^{8 \ga  t} \left(16 \ga ^3-3 D \mu +\ga  \mu  (8 D t+\mu )\right)-2 \mu  e^{4 \ga  t} (\ga  \mu -2 D)\right)},  \label{eq: BM_vel_1_c}
\\
v_2^c(x_1, x_2, t) &=& v_1(x_2, x_1, t) ,
\end{eqnarray}
where the superscript ``$c$" denotes the common environment, and
\begin{eqnarray}
f(x_1, x_2, t) &=& 
e^{8 \ga  t} \left(4 \ga  D \mu  t^2 (3 x_1-x_2)+t (x_1-x_2) \left(16 \ga ^3+\ga  \mu ^2-3 D \mu \right)+4 \ga  D \mu ^3 (x_1+x_2)\right)
\nonumber \\
&+& 2 \mu  e^{4 \ga  t} (\ga  \mu -2 D) \left(2 \ga  t^2 (x_1+x_2)+t (x_2-x_1)+2 \ga  \mu ^2 (x_1+x_2)\right)
\nonumber \\
&+&\mu  (D-\ga  \mu ) \left(4 \ga  t^2 (x_1+x_2)+t (x_2-x_1)+4 \ga  \mu ^2 (x_1+x_2)\right) .
\end{eqnarray}
As expected, in the Schr\"odinger limit $\ga \to 0$ and $D \to 0$, Eq.~\eqref{eq: BM_vel_1_c} reduces to Eq.~\eqref{eq: BM_vel_1_Sch}.

The explicit forms of the quantum forces are too lengthy to be included here; nevertheless, the symmetry condition $F_{\qm, 2}(x_1, x_2, t) = F_{\qm, 1}(x_2, x_1, t)$ still holds. Of particular interest for further analysis is the leading-order correction to the Schrödinger result.
Using the relation $D = 2 \ga T$, the quantum force can be expanded to first order in $\ga$ as
\begin{eqnarray} \label{eq: QM_force_c_approx}
F^c_{\qm, 1}(x_1, x_2, t) & \approx & F^{\Sch}_{\qm, 1}(x_1, x_2, t) + 
\frac{4 \mu (x_1+x_2)t}{\left(\mu ^2 t^2+1\right)^3}
\bigg[ - \mu + \frac{2  \left(\mu ^4 t^4 + 2 \mu ^2 t^2 + 3 \right)}{3} T \bigg ] \ga,
\end{eqnarray}
which correctly reduces to Eq.~\eqref{eq: QM_force_1_Sch} in the Schr\"odinger limit. 
%
%

Starting from the Hamilton-Jacobi-like equation \eqref{eq: HJ-mixed}, one obtains the following Newtonian-like equations of motion
\begin{eqnarray} \label{eq: acc_com}
\frac{d}{dt} v^c_n(x_1, x_2, t) &=& F^c_{\qm,n}(x_1, x_2, t) - 2 \ga (v^c_1 + v^c_2), \qquad n = 1, 2 
\end{eqnarray}
which are consistent with Eq.~\eqref{eq: Newton1}.

For the case of distinct environments, the explicit expressions for the velocity fields are too lengthy to be presented here. However, for the particular case $\mu = 1$, where the particles become decoupled, the velocity fields reduce simplify to
\begin{eqnarray} 
v_n^d(x_n, t) &=& \frac{ 2 \ga ( - 1 + e^{2 \ga t} ) \left( D ( - 1 + e^{2 \ga t} ) + \ga \right)}
{ D \left( - 1 + 4 e^{2\ga t} + e^{4\ga t}  (-3 + 4 \ga t) \right) + \ga \left( 1 - 2 e^{2\ga t} + e^{4\ga t} ( 1 + 4 \ga^2 ) \right)  }  x_n , \qquad n = 1, 2 
\label{eq: BM_vel_n-dis}
\end{eqnarray}
with the superscript ``$d$'' indicating distinct environments.
%
%
%

Consistent with Eq.~\eqref{eq: Newton1}, the corresponding equations of motion take the form
\begin{eqnarray} \label{eq: acc_dis}
\frac{d}{dt} v^d_n(x_1, x_2, t) &=& F^d_{\qm,n}(x_1, x_2, t) - 2 \ga v^d_n(x_1, x_2, t), \qquad n = 1, 2 .
\end{eqnarray}

Finally, in the case of a common environment, the nonlocality measure \eqref{eq: eta} takes the form
\begin{eqnarray} \label{eq: eta_com}
\eta^c(t) &=&
\frac{ |g(t)| }{ \big|
4 \ga \left( \mu^2 + t^2 \right) \left[
\mu  (\ga  \mu -D)+e^{8 \ga  t} \left(16 \ga ^3-3 D \mu +\ga  \mu  (8 D t+\mu )\right)-2 \mu  e^{4 \ga  t} (\ga  \mu -2 D)
\right] \big| } ,
\end{eqnarray}
whereas for distinct environments, it is given by
\begin{eqnarray} \label{eq: eta_dis}
\eta^d(t) &=&
\frac{ \big|
4 \ga^2 \left(\mu^2-1\right) e^{5 \ga  t} \sinh (\ga  t) \left[ 2 \ga  \left(\ga ^2 \left(\mu ^2+1\right)+\ga  D \mu  \left(e^{2 \ga  t}-1\right)+D \mu  t\right)-D \mu  \sinh (2 \ga  t)\right] \big| } { |h(t)| },
\end{eqnarray}
where we have defined 
\begin{eqnarray}
g(t) &=& 
 2 \ga  e^{8 \ga  t} \left[4 \ga  D \mu ^3-4 \ga  D \mu  t^2-t \left(16 \ga ^3+\ga  \mu ^2-3 D \mu \right)\right]+4 \ga  \mu  e^{4 \ga  t} (\ga  \mu -2 D) \left(2 \ga  \mu ^2+2 \ga  t^2+t\right) \nonumber \\
    &~& - \mu  \left[ 2 \ga ^2 \mu  \left(4 \ga  \mu ^2+4 \ga  t^2+t\right)-2 \ga  D \left(4 \ga  \mu ^2+4 \ga  t^2+t\right) \right] , 
\end{eqnarray}
and
\begin{eqnarray}
h(t) &=& \left[ \ga -D \mu +e^{4 \ga  t} \left(4 \ga ^3 \mu ^2+\ga +D \mu  (4 \ga  t-3)\right)-2 e^{2 \ga  t} (\ga -2 D \mu )
\right]  \nonumber \\
    & ~ & \times \left[
\mu  (\ga  \mu -D)+e^{4 \ga  t} \left(4 \ga ^3-3 D \mu +\ga  \mu  (4 D t+\mu )\right)-2 \mu  e^{2 \ga  t} (\ga  \mu -2 D)
\right] .
\end{eqnarray}
As one expects, $ \eta^d(t) $ identically vanishes for $ \mu = 1 $, that is, when the initial state is separable and has no entanglement.

\section{Results and discussion} \label{sec: Res_Dis}

In this section, we present the results of our numerical calculations. As stated earlier, all calculations are performed in dimensionless units, where $m = 1$, $\hbar = 1$, and $k_B = 1$. Alternatively, the results can be interpreted in terms of scaled quantities, where physical variables are rescaled by natural units associated with the system. For instance, positions may be expressed in units of a characteristic length scale such as the initial wavepacket width $\sigma_0$; time in units of $m \sigma_0^2 / \hbar$; the relaxation rate in units of $\hbar / (m \sigma_0^2)$; temperature in units of $\hbar^2 / (m \sigma_0^2 k_B)$; and the density matrix in units of $1 / \sigma_0^2$.
We begin by displaying selected Bohmian trajectories for three different dynamical regimes: unitary evolution governed by the Schrödinger equation, and dissipative dynamics under the CL framework for both distinct and common environmental couplings. We then examine the time evolution of the sensitivity of the Bohmian velocity of one particle to the position of the other as a quantitative measure of Bohmian nonlocality.

\begin{figure} 
\centering
\includegraphics[width=12cm,angle=-0]{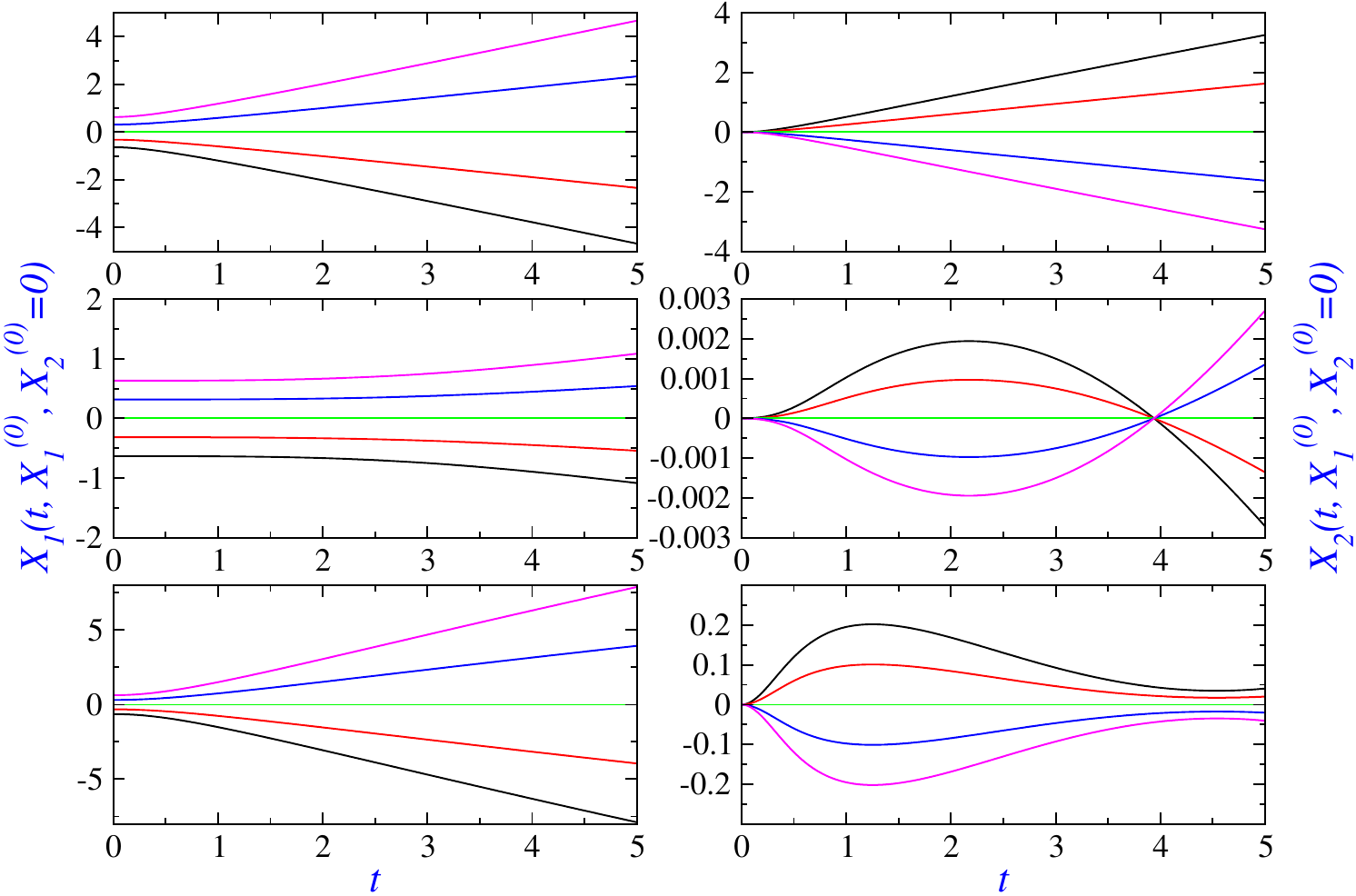}
\caption{
Selected Bohmian trajectories are presented for three dynamical scenarios: unitary evolution under the Schrödinger equation (top panels), dissipative dynamics in the distinct environment case with $\gamma = 0.1$ and $T = 20$ (middle panels), and the common environment case under the same parameters (bottom panels). The left panels display the trajectories of the first particle, while the right panels show those of the second particle. The initial position of the second particle is fixed at $X_2^{(0)} = 0$, and the squeezing decay parameter is set to $\mu = 0.4$.
}
\label{fig: trajs-all} 
\end{figure}

Figure~\ref{fig: trajs-all} shows selected Bohmian trajectories for a bipartite squeezed state under three dynamical scenarios: unitary evolution governed by the Schr\"odinger equation (top panels), and dissipative evolution described by the CL master equation in the cases of distinct environments (middle panels) and a common environment (bottom panels). 
In the dissipative cases, the dissipation rate is set to $\gamma = 0.1$ and the temperature to $T = 20$. The squeezing decay parameter is fixed at $\mu = 0.4$ for all panels. The initial position of particle 2 is set to $X_2^{(0)} = 0$, while $X_1^{(0)}$ is varied across trajectories. The left and right columns display the evolution of particles 1 and 2, respectively.
As expected from the single-valuedness of the wave function in the unitary case, and from the same property reflected in the structure of the density matrix in the CL framework, Bohmian trajectories never intersect in spacetime. A point of potential confusion arises in the middle right panel (distinct-environment case), where some trajectories of particle 2 appear to cross. However, this is not a true intersection. All trajectories of particle 2 are initialized at $X_2^{(0)} = 0$, but they correspond to different values of $X_1^{(0)}$. The apparent crossing thus occurs at the same time but for different spatial locations in the configuration space. The left panel confirms that the position of particle 1 differs at that instant, and hence, the full configuration-space trajectories remain well-separated, consistent with Bohmian theory.
The spread of particle 1’s trajectories (left panels) shows a meaningful dependence on the dynamical regime. Under unitary evolution (top), the trajectories exhibit a noticeable broadening due to free dispersive motion. In the dissipative setting with distinct environments (middle), this broadening is significantly suppressed, reflecting the decohering and frictional effects of local baths. Intriguingly, in the common environment case (bottom), the spread becomes even larger than in the unitary scenario. This counterintuitive result stems from the structure of the CL master equation for a shared bath, which introduces additional cross terms coupling the particles via the environment. These terms can dynamically induce environment-mediated correlations—effectively enhancing the quantum coordination between the particles beyond what is present in the initial state. As a result, even in the presence of decoherence, a common environment can transiently reinforce entanglement-like behavior, leading to a broader dynamical separation than in both the distinct-environment and unitary cases.
The right panels, showing the trajectories of particle 2, all of which are initialized at $X_2^{(0)} = 0$, offer direct insight into nonlocal features of BM. Despite the identical initial condition for particle 2, its evolution depends sensitively on the initial position $X_1^{(0)}$ of particle 1. In the unitary case, for instance, positive (negative) values of $X_1^{(0)}$ lead both particles to move upward (downward) at early times. This directional correlation is a clear signature of nonlocal dependence, rooted in the entangled structure of the wave function and reflected in the quantum potential governing the guiding equations. The same kind of nonlocal sensitivity persists in the dissipative cases, albeit modified by environmental interactions.
At later times, a pronounced bending of the Bohmian trajectories toward the origin is observed in both dissipative cases. This behavior is especially visible in the right panels, which represent the motion of particle 2, all initialized at $X_2^{(0)} = 0$. Since the initial condition is fixed, any variation among these trajectories arises purely from their dependence on the initial position $X_1^{(0)}$, via the conditional density matrix of particle 2. In the Bohmian framework, trajectory dynamics are governed by the conditional wave function (or, in this context, a conditional density matrix \cite{DuGoTuZa-FP-2005}), which evolves under an effective non-unitary equation. When particle 2 is conditioned on the actual position of particle 1, the time evolution of this conditional state determines how sensitive the trajectories of particle 2 remain to variations in $X_1^{(0)}$. In the distinct environment scenario, decoherence progressively reduces this sensitivity, leading to an apparent ``focusing" or narrowing of the spread. In contrast, in the common environment case, the presence of bath-induced correlations preserves a stronger dynamical coupling between the particles. As a result, the trajectories of particle 2 remain more widely separated and do not exhibit the same degree of convergence. Nevertheless, the overall bending back toward the origin---present in both cases---highlights the role of dissipation in suppressing long-range dispersion, regardless of the specific decoherence structure.

\begin{figure} 
\centering
\includegraphics[width=12cm,angle=0]{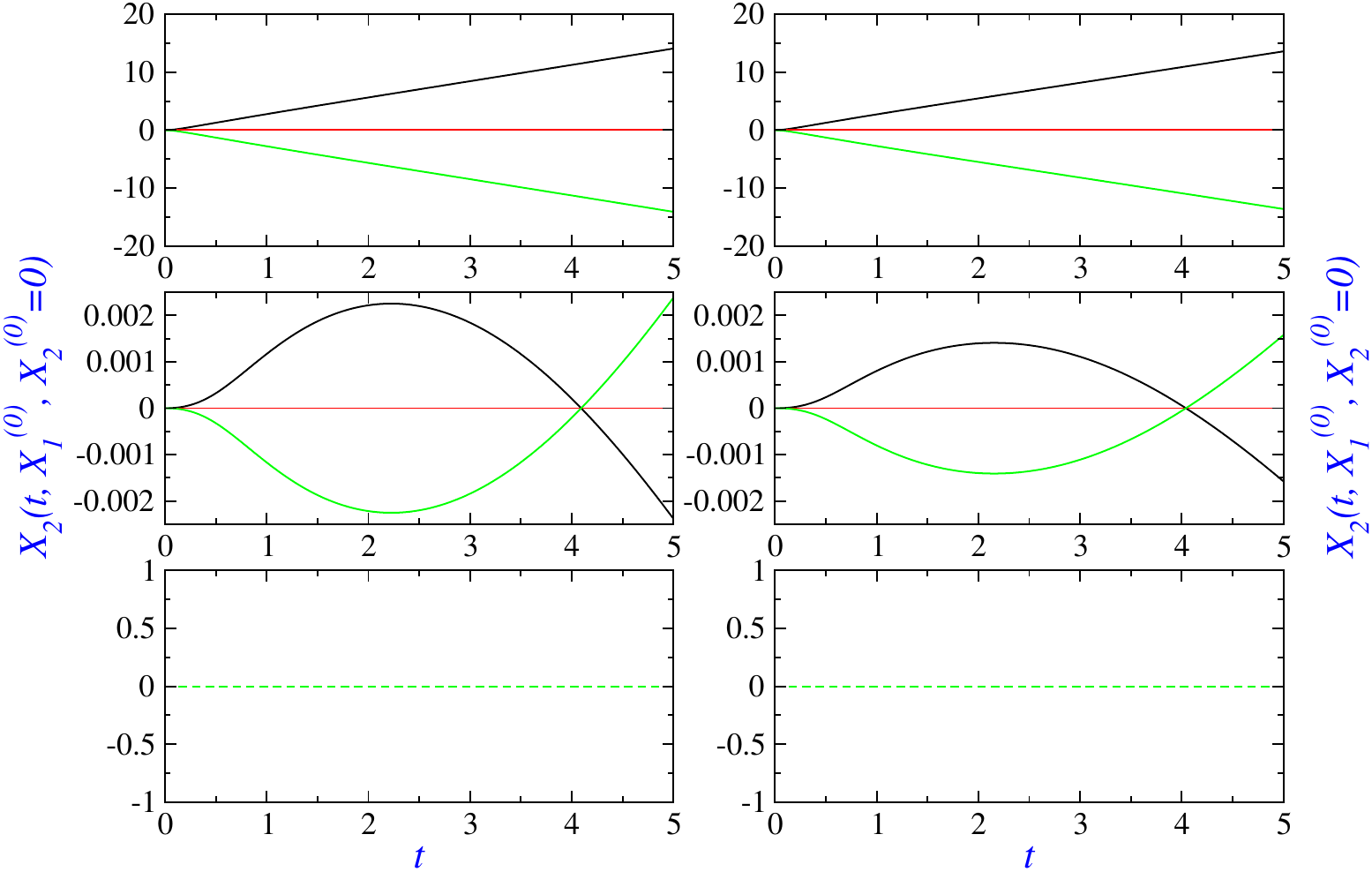}
\caption{
Shown are selected Bohmian trajectories of the second particle within the CL framework for the distinct environments scenario. The left and right panels correspond to temperatures $T = 15$ and $T = 25$, respectively. Each row represents a fixed value of the squeezing decay parameter: $\mu = 0.1$ (top panels), $\mu = 0.5$ (middle panels), and $\mu = 1$ (bottom panels). The initial position of the second particle is fixed at $X_2^{(0)} = 0$, and the relaxation rate is set to $\ga = 0.1$. As expected, for the unsqueezed case ($\mu = 1$), where the particles are initially uncorrelated and independently coupled to distinct environments, the position of the second particle becomes independent of that of the first.
}
\label{fig: trajs-dis} 
\end{figure}

Figure~\ref{fig: trajs-dis} illustrates the Bohmian trajectories of particle 2 under dissipative evolution in the distinct environments scenario, where each particle interacts with its own thermal bath. The trajectories are plotted for two temperatures, $T = 15$ (left column) and $T = 25$ (right column), and for three values of the squeezing decay parameter: $\mu = 0.1$ (top), $\mu = 0.5$ (middle), and $\mu = 1$ (bottom). In all cases, the initial position of particle 2 is fixed at $X_2^{(0)} = 0$, while $X_1^{(0)}$ varies across the trajectories. The relaxation rate is kept constant at $\gamma = 0.1$.
For low values of $\mu$, especially $\mu = 0.1$, the initial entanglement between the particles is strong, and this is clearly reflected in the significant dependence of particle 2's trajectories on the initial position $X_1^{(0)}$ of particle 1. Despite being locally coupled to independent environments, the nonlocal influence of the entangled wave function persists during early times and generates a visible spread in the trajectories of particle 2. This behavior is a direct manifestation of Bohmian nonlocality, as the actual path of particle 2 remains sensitive to the distant initial condition of particle 1, even though both are undergoing decohering dynamics due to their respective baths.
As $\mu$ increases, the initial correlations weaken. For $\mu = 0.5$, this nonlocal dependence is still present but significantly reduced. The spread in particle 2’s trajectories narrows, and the curves begin to cluster more closely together. This indicates that the effective conditional state of particle 2 becomes less sensitive to variations in $X_1^{(0)}$--a signature of partial decoherence and diminishing entanglement.
In the unsqueezed case, $\mu = 1$, the particles are initially in a product state with no entanglement. Since the particles are also coupled to distinct environments, their subsequent evolution remains uncorrelated. As expected, the trajectories of particle 2 show no dependence on $X_1^{(0)}$: all the curves collapse into a single line. This confirms that, in the absence of both initial entanglement and environment-induced correlations (absent here due to the independence of the baths), Bohmian trajectories of particle 2 become completely autonomous. This case serves as a useful reference for assessing the role of quantum correlations in earlier rows.
Regarding the temperature dependence, a modest effect can be observed when comparing the left and right columns. At the higher temperature ($T = 25$), the trajectories in the low-$\mu$ cases are slightly more compressed toward the origin, especially at later times. This reflects the stronger dissipative and decohering influence of the thermal baths at higher temperatures, which tends to suppress the dynamical consequences of the initial entanglement more rapidly. The effect is subtle but consistent, particularly in the top and middle rows, where entanglement initially plays a significant role.
Overall, this figure demonstrates how Bohmian trajectories can provide a clear visualization of nonlocal quantum correlations and their suppression due to environmental decoherence. The interplay between initial entanglement (controlled by $\mu$), thermal noise (via $T$), and environmental structure (distinct vs. common bath) is directly encoded in the dynamical behavior of the trajectories of even a single particle, provided one tracks its evolution across different initial conditions of its entangled partner.

\begin{figure} 
\centering
\includegraphics[width=12cm,angle=-90]{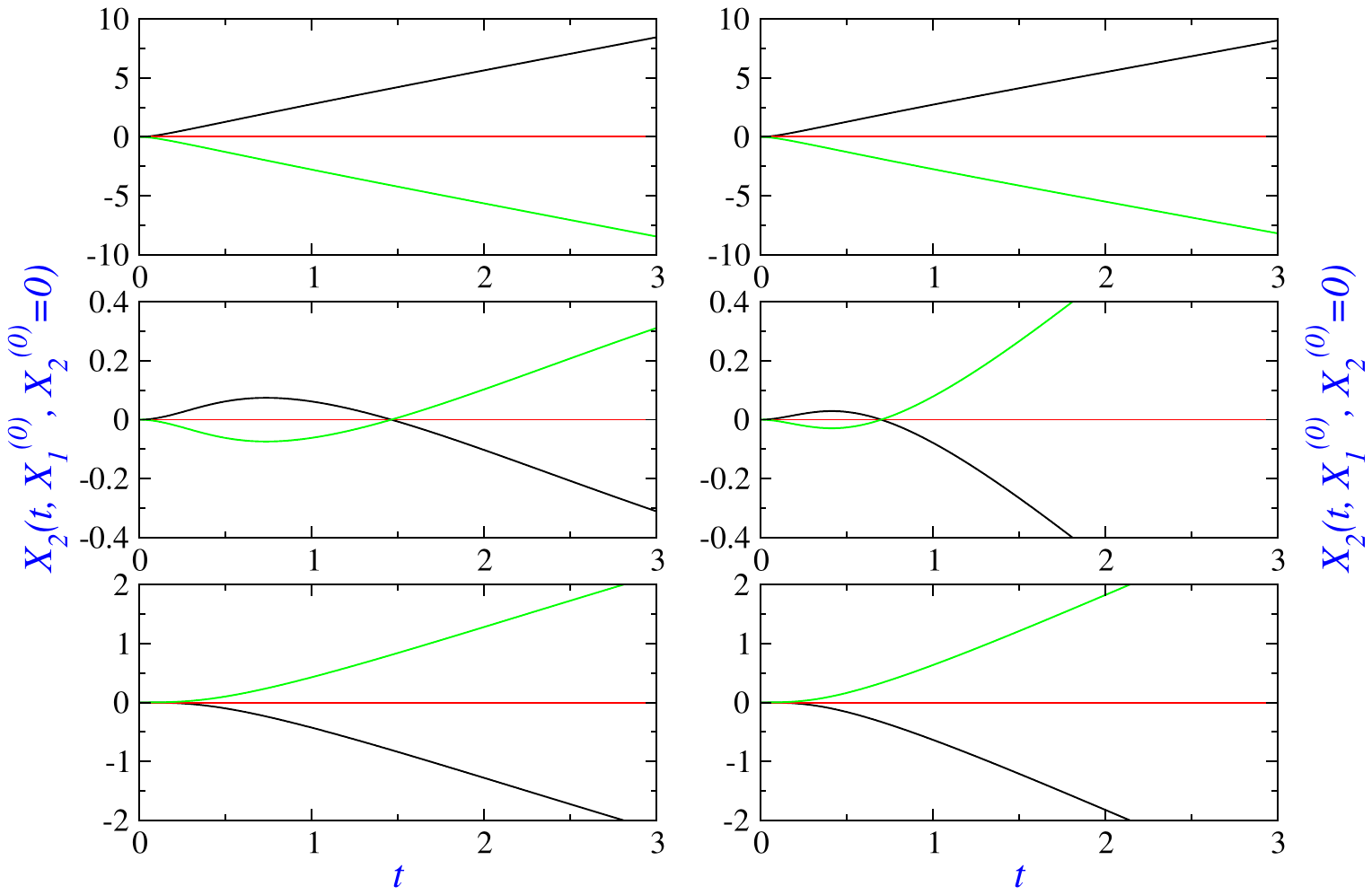}
\caption{
Selected Bohmian trajectories of the second particle within the CL framework for the common environment scenario are shown for $T = 15$ (left panels) and $T = 25$ (right panels). Each row corresponds to a fixed value of the squeezing decay parameter: $\mu = 0.1$ (top panels), $\mu = 0.5$ (middle panels), and $\mu = 1$ (bottom panels). The initial position of the second particle is fixed at $X_2^{(0)} = 0$, and the relaxation rate is set to $\ga = 0.1$.
}
\label{fig: trajs-com} 
\end{figure}

Figure~\ref{fig: trajs-com} displays the Bohmian trajectories of the second particle in the CL framework for the common environment scenario, where both particles are coupled to the same thermal bath. The trajectories are shown for two temperatures, $T = 15$ (left column) and $T = 25$ (right column), and for three values of the squeezing decay parameter: $\mu = 0.1$ (top), $\mu = 0.5$ (middle), and $\mu = 1$ (bottom). In all cases, the initial position of particle 2 is fixed at $X_2^{(0)} = 0$, while $X_1^{(0)}$ varies across the curves. The relaxation rate is set to $\gamma = 0.1$.
%
%
For $\mu = 0.1$, the particles are initially strongly entangled. Given the nonlocal structure of BM, one anticipates that the velocity of particle 2 will exhibit pronounced sensitivity to the initial position of particle 1. This effect is expected to be further enhanced in the presence of a common environment, which may introduce additional correlations through the bath-induced interaction. Such sensitivity manifests as a stronger divergence in the trajectories of particle 2 when the initial position of particle 1 is slightly perturbed.
Surprisingly, however, despite the presence of the shared bath, the overall spread of particle 2's trajectories is found to be smaller in the common environment case than the distinct environment case.
This is evident from the narrower vertical range of trajectories, indicating that the shared environment, under these conditions, acts as a more efficient decohering mechanism, suppressing the manifestation of initial entanglement in the trajectory structure. Furthermore, increasing the temperature from $T = 15$ to $T = 25$ leads to a slight reduction in this spread, consistent with the idea that higher thermal noise accelerates decoherence and thereby reduces the configurational sensitivity of particle 2 to particle 1's initial condition.
As $\mu$ increases to $0.5$, where the initial entanglement is moderate, the behavior changes. The spread of particle 2’s trajectories in the common environment now exceeds that of the distinct environment case. Here, the environment-induced coupling begins to dominate over initial-state features, and the shared bath dynamically sustains or even enhances interparticle correlations. This is further amplified at higher temperatures, where the trajectory spread increases noticeably from left to right. This trend continues into the unsqueezed case $\mu = 1$, where the initial state is fully separable. In the distinct environment case, particle 2's trajectory becomes completely independent of $X_1^{(0)}$, resulting in identical curves. In contrast, under a common bath, particle 2 still shows sensitivity to $X_1^{(0)}$, demonstrating that the bath dynamically induces correlations, even in the absence of initial entanglement.
Temperature plays a nontrivial, context-dependent role. At strong squeezing ($\mu = 0.1$), increasing $T$ suppresses trajectory spread due to enhanced decoherence. However, at weaker squeezing ($\mu = 0.5$ and $1$), increasing temperature leads to greater spread—consistent with correlated thermal fluctuations from the common bath injecting more configurational noise.

It is important to emphasize that the Bohmian trajectories shown here are illustrative and represent a small selection corresponding to specific initial values of $X_1^{(0)}$, with $X_2^{(0)}$ fixed. While they offer valuable qualitative insight into how the second particle's dynamics depend on interparticle correlations and environmental structure, they do not fully characterize the statistical spread of the quantum state. The actual dispersion of the system must be analyzed through ensemble quantities such as the width of the conditional probability distribution or the variance derived from the reduced density matrix. Conclusions drawn from the visual behavior of a few trajectories must therefore be interpreted with care, and always in light of the underlying quantum statistical structure.

\begin{figure} 
\centering
\includegraphics[width=12cm,angle=-0]{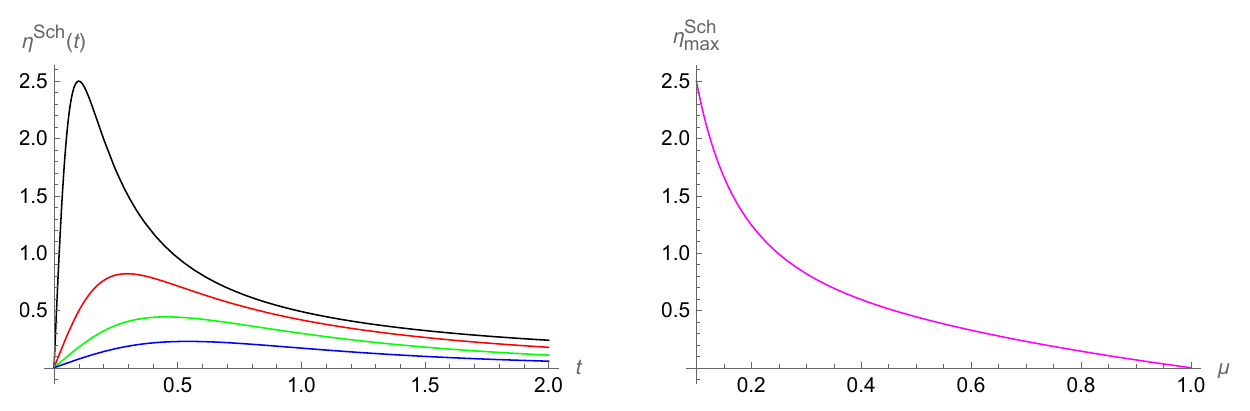}
\caption{
Bohmian nonlocal correlation measure $ \eta^{\Sch}(t) $ for the case of Schr\"odinger has been evaluated for different squeezing decay factors: $ \mu = 0.1 $ (black), $ \mu = 0.3 $ (red), $ \mu = 0.5 $ (green) and  $ \mu = 0.7 $ (blue).
Right panel depicts the maximum value of the Bohmian nonlocal correlation measure in terms of squeezing decay factor.
}
\label{fig: eta_Sch} 
\end{figure}

Figure~\ref{fig: eta_Sch} illustrates the behavior of the Bohmian nonlocal correlation measure $\eta^{\Sch}(t)$ under unitary evolution governed by the Schrödinger equation. The left panel shows the time evolution of $\eta^{\Sch}(t)$, as obtained in Eq.~\eqref{eq: eta_Sch}, for several values of the squeezing decay parameter: $\mu = 0.1$ (black), $\mu = 0.3$ (red), $\mu = 0.5$ (green), and $\mu = 0.7$ (blue). In all cases, the nonlocal measure starts at zero, increases to a peak, and then decays asymptotically. The position and height of this peak depend sensitively on the value of $\mu$. For small $\mu$, corresponding to strong initial squeezing, $\eta^{\Sch}(t)$ exhibits a large and early maximum, reflecting strong nonlocal sensitivity of the Bohmian velocity field. As $\mu$ increases, the maximum decreases in magnitude and shifts to later times, indicating the weakening of nonlocal correlations.
The right panel displays the maximum value $ \eta^{\Sch}_{\max} $, obtained analytically from Eq.~\eqref{eq: eta_Sch_max}, as a function of $\mu$. The observed monotonic decrease confirms that Bohmian nonlocality, as captured by the velocity-based sensitivity to distant positions, is strongly suppressed as the squeezing decay parameter increases. In the limit $\mu \to 1$, where the initial state becomes approximately separable, the measure approaches zero, consistent with the expectation that no nonlocal influence should remain in the Bohmian velocity field. These findings demonstrate that the analytic expression $\eta^{\Sch}(t)$ provides a clear and physically meaningful indicator of entanglement-driven nonlocality in BM.

\begin{figure} 
\centering
\includegraphics[width=12cm,angle=-0]{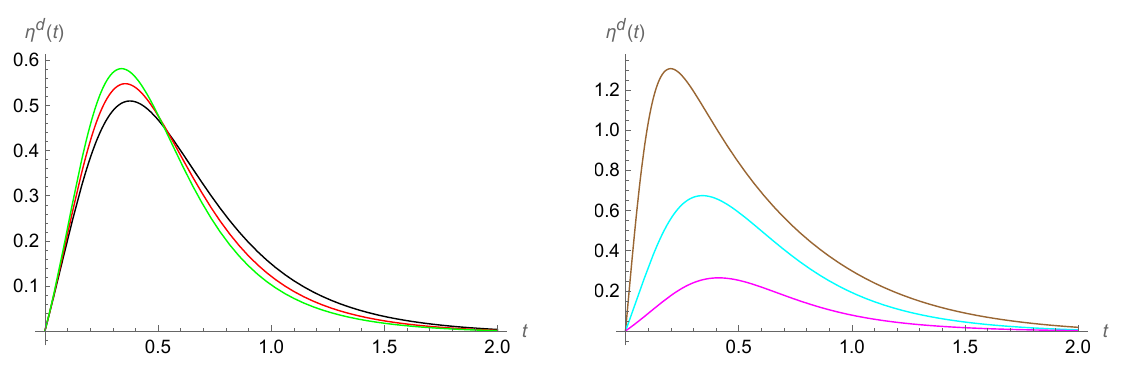}
\caption{
Bohmian nonlocal correlation measure $ \eta^d(t) $ for the case of distinct environments has been evaluated for a relaxation rate of $ \ga=0.1 $.
In the left panel, the squeezing parameter has been fixed at $ \mu = 0.5 $, and different temperatures have been considered: $T = 10$ (black), $T=15$ (red) and $T = 20$ (green). In the right panel, the temperature of both environments has been fixed at $T=10$, while different squeezing decay factors have been used: $ \mu = 0.2 $ (brown), $ \mu = 0.4 $ (cyan) and  $ \mu = 0.7 $ (magenta).
}
\label{fig: eta_dis} 
\end{figure}

The Bohmian nonlocal correlation measure, $ \eta^d(t) $, for distinct environments is examined in figure~\ref{fig: eta_dis} with a fixed relaxation rate of $ \ga = 0.1 $. The figure is divided into two panels, each illustrating the measure's behavior under varying conditions. The left panel illustrates the effect of temperature on the Bohmian nonlocal correlation measure $ \eta^d(t) $ for a fixed squeezing parameter. In all cases, the measure exhibits a transient structure characterized by an initial increase from zero, a well-defined peak at intermediate times, and a subsequent decay toward zero. As temperature increases, the peak value of $ \eta^d(t) $ rises, reflecting a stronger instantaneous nonlocal response due to enhanced thermal fluctuations. However, this increase is accompanied by a steeper decline following the peak, such that the measure for higher temperatures eventually drops below that of lower temperatures. A clear manifestation of this effect is seen in the full width at half maximum (FWHM) of each curve: as temperature increases, the FWHM systematically decreases; it is 0.6737 for $ T = 10 $, 0.6072 for $ T = 15 $ and 0.5612 for $ T = 20 $. This indicates that although higher temperatures transiently enhance the nonlocal correlation, the duration over which it remains significantly strong becomes shorter. At later times, all curves decay to zero and converge around $t \approx 2$, indicating that environmental decoherence ultimately suppresses nonlocal effects regardless of temperature. These observations demonstrate that temperature modulates both the amplitude and the temporal width of Bohmian nonlocal correlations in open quantum systems.
The right panel examines the effect of the squeezing decay parameter $\mu$ at fixed temperature. Similar to the temperature dependence, decreasing $\mu$ enhances both the peak height and the early-time emergence of the nonlocal correlation measure. However, unlike the case of temperature, the FWHM values do not exhibit a simple monotonic trend with $\mu$. Specifically, the FWHM is 0.5727 for $ \mu = 0.2 $, 0.6721 for $\mu = 0.4$, 0.6680 for $\mu = 0.7$, and 0.6644 for $\mu = 0.9$. These values indicate that the temporal width of the correlation is smallest for the strongest squeezing (i.e., $\mu = 0.1$), but increases for moderate values of $\mu$ and then remains nearly constant as $\mu$ becomes larger. This behavior suggests that, beyond an initial sensitivity to squeezing, the correlation duration becomes largely insensitive to further increases in $\mu$. Taken together, both panels demonstrate that temperature and squeezing play distinct but complementary roles in shaping Bohmian nonlocal correlations: temperature modulates both the strength and the temporal width in a correlated fashion, while squeezing predominantly governs the amplitude and onset time, with a more subtle influence on the lifetime of the correlations.

\begin{figure} 
\centering
\includegraphics[width=12cm,angle=-0]{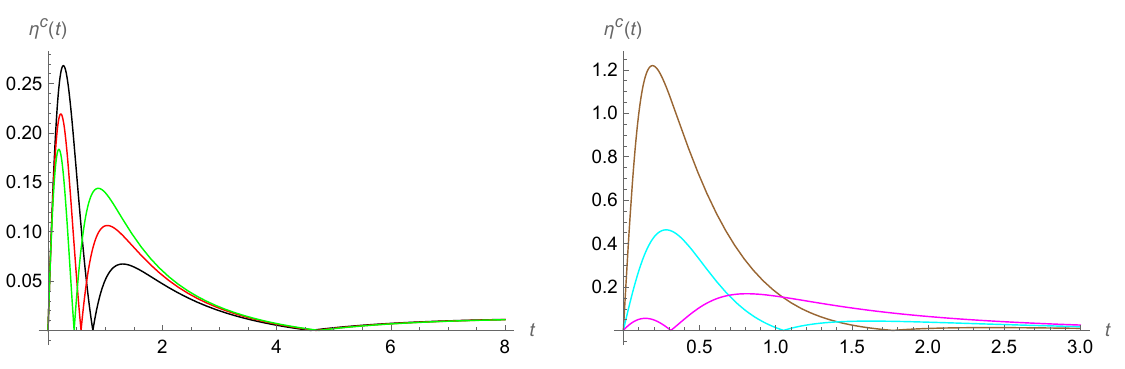}
\caption{
Bohmian nonlocal correlation measure $ \eta^c(t) $ for the case of common environment has been evaluated for a relaxation rate of $ \ga=0.1 $.
In the left panel, the squeezing parameter has been fixed at $ \mu = 0.5 $, and different temperatures have been considered: $T = 10$ (black), $T=15$ (red) and $T = 20$ (green). In the right panel, the temperature of the common environment has been fixed at $T=10$, while different squeezing decay factors have been used: $ \mu = 0.2 $ (brown), $ \mu = 0.4 $ (cyan) and  $ \mu = 0.7 $ (magenta).
}
\label{fig: eta_com} 
\end{figure}

The Bohmian nonlocal correlation measure, $ \eta^c(t) $, for a common environment scenario is evaluated with a relaxation rate of $\gamma = 0.1$ in figure \ref{fig: eta_com}. This figure also consists of two panels, examining the measure's behavior under different conditions, and offers interesting comparisons with the distinct environment case.
In the left panel, the squeezing parameter is fixed at $ \mu = 0.5 $, and various temperatures are considered: $T=10$  (black), $T=15$  (red), and $T=20$  (green). In contrast to $ \eta^d(t) $, where the correlation showed a single peak and smooth decay, $ \eta^c(t) $ exhibits oscillatory behavior with multiple peaks and dips, indicating a more complex temporal evolution of nonlocality in a common environment. The initial increase and peak are still present, but subsequent oscillations suggest a dynamic interplay of factors. 
%
Increasing the bath temperature reduces both the height and width of the initial peak in the sensitivity measure, reflecting stronger short-time decoherence. However, the first revival becomes more pronounced and longer-lasting at higher temperatures, indicating that thermal fluctuations can transiently enhance the effective nonlocal correlations induced by the common bath. This counter-intuitive behavior highlights the subtle role of temperature in the interplay between decoherence and bath-induced correlations.
The right panel investigates the influence of the squeezing decay factor on $ \eta^c(t) $ , with the temperature of the common environment fixed at $T=10$. Different squeezing decay factors are used: $\mu = 0.2$ (brown), $\mu = 0.4$ (cyan), and $\mu = 0.7$ (magenta). Similar to the distinct environment case, a smaller squeezing decay factor ($\mu = 0.2$) results in a significantly higher initial peak in $ \eta^c(t) $ compared to larger $\mu$ values. The brown curve ($\mu = 0.2$) shows the most prominent initial peak, followed by the cyan curve ($\mu = 0.4$), and the magenta curve ($\mu = 0.7$) exhibits the lowest initial peak. This reinforces the inverse relationship between the squeezing parameter and the strength of the initial nonlocal correlation observed in the distinct environment case. However, the oscillatory behavior is also evident here, particularly for smaller $\mu$ values, indicating that the common environment introduces a more complex dynamic regardless of the squeezing parameter. The decay of correlations over time is still present, but the oscillations suggest that nonlocality might temporarily reappear or strengthen at later times, unlike the smoother decay seen in $ \eta^d(t) $.
In comparison to the previous figure for distinct environments, the most striking difference is the oscillatory nature of $ \eta^c(t) $ versus the single-peak, smooth decay of $ \eta^d(t) $. This suggests that a common environment introduces a more complex and possibly recurrent manifestation of nonlocal correlations. While both scenarios show that smaller squeezing parameters generally lead to stronger initial nonlocal correlations, the common environment introduces a richer temporal dynamic characterized by oscillations and potentially multiple local maxima.

A comparison between the two scenarios underscores the decisive role of the environment in shaping quantum dynamics. In both cases, the Bohmian nonlocal correlation measure initially increases from zero, reflecting the natural build-up of correlations before environmental influences dominate. However, the subsequent evolution differs fundamentally. For distinct environments, the correlation measure undergoes a monotonic decay after reaching its peak. In contrast, the common environment induces a non-monotonic decay marked by oscillations and partial revivals of the nonlocal measure.
The observed revivals and oscillations in the case of a common reservoir should not be taken as definitive evidence of non-Markovian behavior. The underlying dynamics are governed by the two-particle CL master equation in the high-temperature and weak-coupling limit---conditions under which Markovian approximations are typically valid. A rigorous assessment of non-Markovianity would require the application of standard criteria, such as those based on trace distance \cite{BrLaPi-PRL-2009} or complete positivity divisibility \cite{ReHuPl-PRL-2010}, which are computationally infeasible here due to the infinite-dimensional Hilbert space and the complexity of the density matrix. 
Additionally, the role of temperature is reversed: whereas higher temperatures enhance nonlocal correlations in the distinct case, they suppress them in the common environment. These observations demonstrate that the structure of the system-environment coupling governs not only the quantitative aspects of decoherence such as correlation strength and lifetime, but also the qualitative nature of the dynamics.
Finally, we note that entanglement generation via a common but Markovian environment has been previously reported in the literature \cite{BeFlPi-PRL-2003}.

\section{Summary and conclusions} \label{sec: Sum_con}

This work has investigated the influence of different environmental structures on the dynamics of quantum correlations in bipartite open quantum systems within the Bohmian framework.
Our results emphasize the crucial role of the environmental structure in shaping the dynamics of quantum correlations. When the two particles interact with distinct, independent environments, the nonlocal correlations exhibit a smooth, monotonic decay consistent with Markovian dissipation. In contrast, a common environment induces more complex temporal behaviors characterized by revivals and oscillations in the measure of nonlocality. Crucially, these behaviors manifest within the Markovian framework of the CL equation, arising from the complexity system-environment coupling rather than genuine memory effects. We further observe that temperature influences the two scenarios differently: increasing temperature tends to enhance the peak magnitude of correlations in distinct, independent environments while narrowing their temporal duration, whereas in a common environment, temperature modulates the amplitude and timing of the revivals. These findings demonstrate that the structure of system-environment coupling is a fundamental determinant of both the qualitative and quantitative aspects of quantum decoherence and correlation dynamics.
Our study shows that the fate of quantum correlations in open bipartite systems is governed by a subtle interplay between initial entanglement, environmental dissipation, and the structure of system-bath interactions. While local, independent environments cause a gradual and monotonic decay of coherence and nonlocal correlations, a shared bath induces effective coupling that leads to transient revivals and oscillations in these correlations despite dissipation. 
Interestingly, while increasing temperature suppresses the initial peak of the Bohmian sensitivity measure, it enhances the amplitude and duration of the first revival, highlighting the nontrivial role of thermal fluctuations in transiently amplifying bath-induced nonlocal effects.
By employing Bohmian trajectories and associated nonlocal correlation measures, we provided both qualitative and quantitative insight into the dynamics of entanglement and decoherence. Although the trajectory approach offers a vivid and intuitive picture of quantum nonlocality's evolution, it must be complemented by statistical ensemble analyses to fully capture the underlying quantum state dynamics.
In conclusion, understanding and controlling the environmental structure and parameters is essential for manipulating quantum correlations in realistic noisy systems. These findings open promising avenues for exploring environment-assisted preservation and engineering of quantum coherence, with implications for quantum technologies subject to decoherence. Moreover, by adopting the BM framework, this study provides an intuitive, trajectory-based perspective on nonlocality and decoherence dynamics, enriching our conceptual and practical understanding of open quantum systems.

\vspace{2cm}

\noindent
{\bf Acknowledgement}:
Support from the university of Qom is acknowledged. We express our gratitude to the referees and editor for their thorough review and constructive suggestions. The author acknowledges the use of generative-AI tools for language editing and takes full responsibility for the content of the manuscript.

\vspace{0.5cm}
\noindent
{\bf Data availability}: This manuscript has no associated data.



%
\end{document}